\documentclass[a4paper,12pt]{article}

\usepackage[center]{titlesec}

\renewcommand{\thesection}{\Roman{section}}

\renewcommand{\thesubsection}{\arabic{subsection}}

\renewcommand{\thesubsubsection}{\alph{subsubsection}}

\titleformat%
    {\section}
    [block]
    {\centering\bfseries}
    {\thesection.}
    {0.5em}
    {}

\titleformat%
    {\subsection}
    [block]
    {\centering\itshape}
    {\thesubsection.}
    {0.5em}
    {}

\titleformat%
    {\subsubsection}
    [block]
    {\centering}
    {\thesubsubsection.)}
    {0.5em}
    {}

\usepackage{lipsum}

\titlespacing{\section}{0pt}{*4}{*1.5}
\titlespacing{\subsection}{0pt}{*4}{*1.5}
\titlespacing{\subsubsection}{0pt}{*4}{*1.5}

\usepackage[utf8]{inputenc}
\usepackage{cancel}
\usepackage{tikz}
\usepackage{ulem}
\usepackage{amsfonts}
\usepackage{amssymb}
\usepackage{graphicx}
\usepackage{amsmath}
\usepackage{enumerate}
\usepackage{mathtools}
\usepackage{subfig}
\usepackage{color}
\usepackage{tikz}
\usepackage{float}
\usepackage{here}
\usepackage{cite}
\usepackage{mathrsfs}
\usepackage{float,epsfig}
\usepackage{dcolumn}
\usepackage{graphicx}
\usepackage{bm}
\usepackage{amsmath,amssymb,amsthm}
\usepackage[colorlinks=true,linkcolor=blue,citecolor=red]{hyperref}
\usepackage{multirow}
\usepackage[toc,page]{appendix}
\usetikzlibrary{arrows.meta}
\usetikzlibrary{bending}
\usetikzlibrary{calc}
\newcommand{\be}{\begin{equation}}
\newcommand{\ee}{\end{equation}}
\newcommand{\bea}{\setlength\arraycolsep{2pt} \begin{eqnarray}}
\newcommand{\eea}{\end{eqnarray}}

\def\0{{\sst{(0)}}}
\def\1{{\sst{(1)}}}
\def\2{{\sst{(2)}}}
\def\3{{\sst{(3)}}}
\def\4{{\sst{(4)}}}
\def\5{{\sst{(5)}}}
\def\6{{\sst{(6)}}}
\def\7{{\sst{(7)}}}
\def\8{{\sst{(8)}}}
\def\sst#1{{\scriptscriptstyle #1}}

\makeatletter \@addtoreset{equation}{section}

\definecolor{lime}{HTML}{A6CE39}

\begin{document}

\title{{\normalsize \textbf{\Large    Swampland Program for   Hypergeometric Inflation Scenarios in  Rescaled Gravity }}}
\author{ {\small  Saad Eddine Baddis\footnote{saadeddine.baddis@um5r.ac.ma}    \;  and  Adil  Belhaj\footnote{a-belhaj@um5r.ac.ma}    \thanks{%
\bf Authors in alphabetical order.} \hspace*{-8pt}} \\
{\small D\'{e}partement de Physique, \'Equipe des Sciences de la
mati\`ere et du rayonnement, ESMaR} \\ {\small Facult\'e des Sciences, Universit\'e Mohammed V de Rabat, Rabat,
Morocco}}
\maketitle

\begin{abstract}
In this paper, we investigate   hypergeometric  stringy  corrections in the swampland program  for    rescaled gravity.   Precisely,  we study   inflationary  models from   Gauss-Bonnet  hypergeometric scalar couplings via  the falsification scenario.  We first    derive   generalized exponential potentials from such hypergeometric behaviors.   Then, we examine certain  selected scalar potentials by  computing the relevant  cosmological quantities  using   the slow-roll mechanism.  Choosing  specific points in the  corresponding moduli space, we provide viable findings  corroborated by  Planck  observational  data and checked by  the swampland criteria. 

\textbf{Key words}: Rescaled Einstein-Hilbert gravity, Inflation,   Slow-roll mechanism, Swampland Conjectures, Hypergeometric  functions.
\end{abstract}


\section{Introduction}
Recently, the swampland program has received a remarkable interest in connection with various physical theories  including the dark dimension \cite{D1,D2,D3}. Precisely, this  scenario has been explored to bridge different dark sectors via the compactification mechanism \cite{IST1,IST2,IST3}. Certain predictions have been obtained which could be corroborated by  empirical investigations by means of  the falsification scenario. Alternatively, the swampland program has been  combined with inflationary building model activities  where the scalar potential plays a relevant role.   This scalar  potential   depends on  the scalar fields.  The most studied  models  in such a program involve only one scalar field identified with the inflaton \cite{IST4,IST44,IST444,IST04,IST004,IST0004,IST00004}.  
To validate the obtained  inflationary models, many constraints  have been  imposed on such a scalar quantity. Slow-roll approximations, for instance, have been  exploited  to provide  numerical values of primordial inflationary observables belonging to ranges accepted by experimental findings \cite{IST5,IST6,IST7,IST8,IST9,IST10,IST11,IST12}.  In this context, various models have been developed and examined  including the ones extracted from brane physics and string theory using the compactification  mechanism \cite{IST1,IST2,IST3,IST30,IST300,IST33,IST333}. Moreover,  the swampland criteria have been approached for modified gravities  to generate corroborated inflationary models.   It is recalled that   such modified gravity theories   have been largely investigated   in connection with  inflationary  building models and black hole physics \cite{I0,I00,I1,I2,I3,I4,I44,I5,BH1,BH2}.  A close examination shows that  the  most dealt with  ones are $F(R)$  gravity actions 
 where $R$  is  the Ricci scalar   assembled  with  various scalar potentials\cite{r23,r24,r25,r26,r27}.  Using  the 
slow-roll  mechanism,  the  spectral index $n_s$ and the tensor/scalar
ratio $r$ have been computed  supplying  findings  being in agreement with   the observational  data \cite{IST5,IST6,IST7,IST8,IST9}.  In order to make contact with dark sectors, extended modified gravity models have been also  studied by implementing extra quantities  like  the trace  of
the stress-energy tensor \cite{adil}.

More recently,  a special emphasis has been on  a rescaled gravity  with  $F(R)=\alpha R$, where $\alpha$  is a  relevant parameter,   intriguing  correction   terms animated by string theory spectrum  \cite{RG1,RG2,RG3,RG4,RG5,RG6,RG7}.   These stringy  corrections have been added by means of  the Gauss-Bonnet  (GB) term with a coupling  function denoted by  $\xi$.  For certain models, this  function solely depends  on   one  scalar field $\phi$ needed to measure the contribution of the stringy  correction. This  scenario   has been  explored and exploited to bring on    scalar potential forms matching with the swampland conjecture.
\par
The aim of this   paper  is to model  hypergeometric  stringy  corrections  in  the swampland program for  $F(R)=\alpha R$  rescaled gravity. Concretely,  we   investigate  inflationary  models from such hypergeometric  coupling behaviors via the  falsification scenario.   We  first derive  generalized exponential  scalar potentials from such hypergeometric coupling functions.   Then, we  approach   and examine certain  selected  potentials by determining  the relevant cosmological quantities via    the slow-roll mechanism.  Choosing  specific points in the  involved  moduli space, we provide viable findings  corroborated by  Planck  observational  data and proven by  the swampland criteria.

This work  is  structured as follows: In section II,  we present  the swampland criteria for  GB stringy corrections in the   $F(R)=\alpha R$ rescaled gravity.   In section III,  we  propose  and discuss  inflationary  models from hypergeometric scalar coupling scenarios.  In section IV,  we  derive generalized exponential potentials from  such coupling  behaviors in the   swampland program combined with the falsification mechanism.  The last section concerns  concluding remarks. 
\section{Swampland criteria for GB stringy corrections in rescaled gravity}
In this section, we give    a  succinct  discussion on  the swampland criteria for  GB stringy corrections in  the  rescaled gravity with  $F(R)=\alpha R$. The latter has been  studied  in detail in many works \cite{RG1,RG2,RG3,RG4,RG5,RG6,RG7}.  Roughly,  we consider the following action with a  GB stringy correction
\begin{equation}
\label{e0021}
S=\int d^4x\sqrt{-g}\left[\frac{\alpha }{2\kappa^2}R-\frac{1}{2}\partial_{\mu}\phi\partial^{\mu}\phi-V(\phi)
-\xi(\phi){\cal G}\right]
\end{equation}
where one has used  $M_p^{-2}=8\pi G=1$. $\alpha$ is a rescaled parameter conditioned by $0<\alpha<1$.  The  fundamental pieces  of this action are the scalar field $\phi$ and the scalar potential $V(\phi)$. $G_{\mu\nu}$ and $\cal G$ denote the Einstein tensor  and the GB term being given by ${\cal G}=R^2-4R^{\mu\nu}R_{\mu\nu}+R_{\mu\nu\lambda}R^{\mu\nu\lambda}$, respectively. The scalar coupling  function $\xi(\phi)$ is a relevant quantity in the present investigation which describes the stringy correction to the rescaled gravity.  Throughout this paper, we  will see that this quantity could provide certain scalar potential forms satisfying the swampland criteria.  Using the Freedman-Robertson-Walker-Lemaitre metric $ds^2=dt^2-a(t)(dx^2+dy^2+dz^2)$ where $a(t)$ is the scalar factor, one can obtain the equations of motion. Precisely,  they are expressed as follows 
\begin{eqnarray}
\frac{3\alpha H^2}{\kappa^2} &=& \frac{1}{2}\dot{\phi}^2+V(\phi)+24\dot{\xi}H^3,\\
-\frac{2\alpha \dot{H}}{\kappa^2} &=& \dot{\phi}^2-16\dot{\xi}H\dot{H}-8H^2(\ddot{\xi}-H\dot{\xi}),\\
\ddot{\phi}+3H\dot{\phi}+V^\prime(\phi)+\xi^\prime\mathcal{G} &=& 0
\end{eqnarray}
where the prime denotes the derivative with respect to $\phi$ and the dot is the time derivative. $H$ is the Hubble parameter given by $H=\frac{\dot{a}}{a}$. These equations can be simplified using the slow roll conditions. Indeed, they  can be reduced as follows
 \begin{eqnarray}
 \label{eqdif}
\frac{3\alpha H^2}{\kappa^2} &\simeq & V(\phi),\\
-\frac{2\alpha \dot{H}}{\kappa^2} &\simeq & \big(\frac{H\xi^\prime}{\xi^{\prime\prime}}\big)^2,\\
V^\prime (\phi)+\frac{\xi^\prime}{\xi^{\prime\prime}}\frac{\kappa^2V(\phi)}{\alpha}+\frac{8}{3\alpha^2}\xi^\prime\kappa^4V(\phi)^2  &\simeq & 0.
\end{eqnarray}
The crucial cosmological observables can be obtained. Concretely, they  reads as 
\begin{eqnarray}
n_S &\simeq  & 1-2(\epsilon_1+\epsilon_2+\epsilon_3),\\
n_\tau &\simeq & -2(\epsilon_1+\epsilon_4),\\
r &\simeq & \vert 16\left(\alpha\epsilon_1-\frac{\kappa^2Q_e}{4H}\right)\frac{c^3_\mathcal{A}}{\kappa^2Q_{GB}}\vert.
\end{eqnarray}
In this way, the slow-roll indices $\epsilon_i$ are  given by
\begin{eqnarray}
\label{Eq3.1}
\epsilon_1 &=& -\frac{\dot{H}}{H^2},\qquad \epsilon_2 = \frac{\ddot{\phi}}{H\dot{\phi}},\\
 \label{Eq3.2}
\epsilon_3 &=& \frac{\dot{E}}{2HE},\qquad \epsilon_4 = \frac{\dot{Q}_{GB}}{2HQ_{GB}}.
\end{eqnarray}
The  involved auxiliary functions  take the following forms
\begin{eqnarray}
E &=& \frac{\alpha}{\kappa^2}(1+\frac{3Q^2_a}{2Q_{GB}\dot{\phi}^2}) \\ {Q}_{GB} &=& \frac{\alpha}{\kappa^2}-8\dot{\xi}H
\end{eqnarray}
where one has used  $ Q_a = -8\dot{\xi}H^2$,  $  Q_f = 8(\ddot{\xi}-H\dot{\xi})$ and  $Q_e = -32\dot{\xi}\dot{H}$. The number of  e-foldings is 
\begin{equation}
N=\int^{t_f}_{t_i}Hdt=\int^{\varphi_f}_{\phi_i}\frac{H}{\dot{\phi}}d\phi=\int^{\phi_f}_{\varphi_i}\frac{\xi^{\prime\prime}}{\xi^\prime}d\phi
\end{equation}
where $\phi_i$ and $\phi_f$ are the initial and  the final values of the scalar field, respectively. It has been remarked  that the ratio $\frac{\xi^\prime}{\xi^{\prime\prime}}$ is vital  in the derivation of acceptable inflationary  models. The  appearance of this ratio in many expressions gives the importance of the string theory  correction function $\xi(\phi)$ in  such investigation directions. In the present work, we  reveal that   this  function  could generate certain scalar  potentials corroborated by observational data. In fact,  extra constraints on  such potential forms could  be imposed via the swampland program. The latter has been introduced in \cite{R1,R11,R2,R3,R4,R5} to check for the consistency with quantum gravity  in the high energy regime.  Roughly, the swampland criteria are usually: 
\begin{itemize}
\item the swampland distance conjecture being
\begin{equation}
\mid \kappa\Delta\phi\mid<\mathcal{O}(1),
\end{equation}
\item the de sitter conjectures which  are 
\begin{eqnarray}
\frac{\mid V^\prime(\phi)\mid}{\kappa V(\phi)}>\mathcal{O}(1), \qquad 
-\frac{V^{\prime\prime}(\phi )}{\kappa^2V(\phi)}>\mathcal{O}(1).
\end{eqnarray}
\end{itemize} 
Having brought  the essential materials, we move to investigate  inflationary models from  hypergeometric  coupling scenarios  for certain regions of the associated moduli space.
\section{Inflationary  models from hypergeometric scalar couplings}
To start, we propose the following stringy  coupling function 
\begin{equation}
\xi(\phi)=\lambda\int^{\kappa\phi}_0 e^{-\frac{\beta}{m}\lbrace\frac{Csc(\frac{m-2}{n}\pi)}{n}+\frac{x^{2-(m+n)}}{m+n-2}\ _2F_1(1,\frac{m+n-2}{n},\frac{m+2n-2}{n},-x^{-n})\rbrace}dx+\eta
\end{equation}
being a solution of the  differential  equation \begin{equation} \frac{\xi^\prime(\phi)}{\xi^{\prime\prime}(\phi)} = -\frac{m(\kappa\phi)^{m-1}((\kappa\phi)^n+1)}{\beta}.  \end{equation}
 $\lambda$ and $\eta$ are integration constants. At this level,  we could provide comments on the parameters. First,   the integration constant $\eta$ does  not have any significance on the relevant quantities. Thus, it can be  considered  as a residual associated with the theory.   The second comment  concerns the   parameter  $m$. Using the involved functions,  we can derive   constraints on  $m$  for the   proposed coupling scenario.  For  $x>0$, it  is conditioned by  $m<2$. For simplicity reasons, we consider  only  $m=1$.  The general  statement  is beyond the scope of
the present work. It  could be approached   in future works. In this way,    the investigated  models   will be labeled only by the parameter $n$. It has been remarked that the above   coupling function is not normalizable, which seems to be a characteristic of all polynomial scalar potentials which could  be  implemented in the action  given by Eq.(\ref{e0021}). Solving the differential equation appearing in    Eq.(\ref{eqdif}), we obtain the scalar potential
\begin{equation}
V(\phi)=\frac{e^{-\frac{\kappa^{3}}{\beta\alpha}\phi(\frac{1}{1+n}(\kappa\phi)^n+1)}}{c+\frac{8\kappa^4}{3\alpha^2}\int^{\kappa\phi}_0 \xi^\prime(x) e^{-\frac{\kappa ^3x}{\beta}(\frac{(\kappa x)^n}{1+n}+1)}dx}
\end{equation}
where $c$ is an  integration constant.  After calculations,  we  get the number of the  e-foldings 
\begin{equation}
N=\beta\left(\phi_f\ _2F_1(1,\frac{1}{n};\frac{1+n}{n};\phi_f^n)-\phi_i\ _2F_1(1,\frac{1}{n};\frac{1+n}{n};\phi_i^n)\right).
\label{Eq.3.4}
\end{equation}
For convenience,  we set $\kappa=1$. In this way,   the first two slow roll indices are
\begin{eqnarray}
\epsilon_1 &=& \frac{1}{2\alpha\beta^2}(\phi^n+1)^2\\
\epsilon_2 &=& 1-\frac{\xi^\prime(\phi)}{\xi^{\prime\prime}(\phi)}\frac{\xi^{\prime\prime\prime}(\phi)}{\xi^{\prime\prime}(\phi)}-\epsilon_1
\end{eqnarray}
where one  has found 
\begin{eqnarray}
\frac{\xi^{\prime\prime\prime}(\phi)}{\xi^{\prime\prime}(\phi)} &=& \frac{n\ _2F_1(1,\frac{n-1}{n},\frac{2n-1}{n},-\phi^{-n})}{\phi\big(\ _2F_1(1,\frac{n-1}{n},\frac{2n-1}{n},-\phi^{-n})+n\phi\ _2F_1(2,\frac{2n-1}{n},\frac{3n-1}{n},-\phi^{-n})\big)}\notag\\&+&\frac{n\frac{1+3n}{2n-1}\ _2F_1(2,\frac{2n-1}{n},\frac{3n-1}{n},-\phi^{-n})}{\phi^{1+n}\big(\ _2F_1(1,\frac{n-1}{n},\frac{2n-1}{n},-\phi^{-n})+n\phi\ _2F_1(2,\frac{2n-1}{n},\frac{3n-1}{n},-\phi^{-n})\big)}\notag\\&+& \frac{n^2\frac{n-1}{3n-1}\ _2F_1(3,\frac{3n-1}{n},\frac{4n-1}{n},-\phi^{-n})}{\phi^{1+2n}\big(\ _2F_1(1,\frac{n-1}{n},\frac{2n-1}{n},-\phi^{-n})+n\phi\ _2F_1(2,\frac{2n-1}{n},\frac{3n-1}{n},-\phi^{-n})\big)}\notag\\
&+& \frac{\xi^{\prime\prime}(\phi)}{\xi^\prime(\phi)}.\end{eqnarray}
 The field values can be obtained from the  final state of inflation required by  $\epsilon_1(\phi_f)=1$.  Indeed,  such values should satisfy  the following equation
\begin{equation}
\phi_f=\left(\beta\sqrt{2\alpha}-1\right)^{\frac{1}{n}}.
\label{Eq3.3}
\end{equation}
In the following section, we exploit a numerical approach to illustrate such an investigation by considering specific models labeled   by the parameter $n$.
\section{ On generalized exponential potentials}
In this section,  we would like to  expose a detailed discussion on certain selected $n$-models. Precisely, we give the obtained numerical values of the involved quantities including the ones associated with the swampland criteria.
\subsection{Model with $n=2$}
\begin{figure}[!ht]
\begin{center}
\centering
\begin{tabbing}
\centering
\hspace{8.cm}\=\kill
\includegraphics[width=8cm, height=7cm]{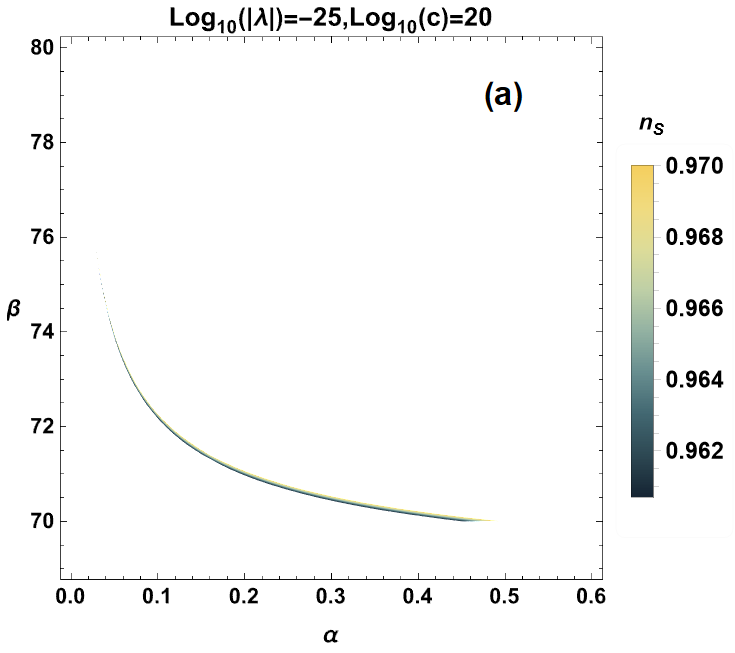}
\hspace{0.1cm} \includegraphics[width=8cm, height=7cm]{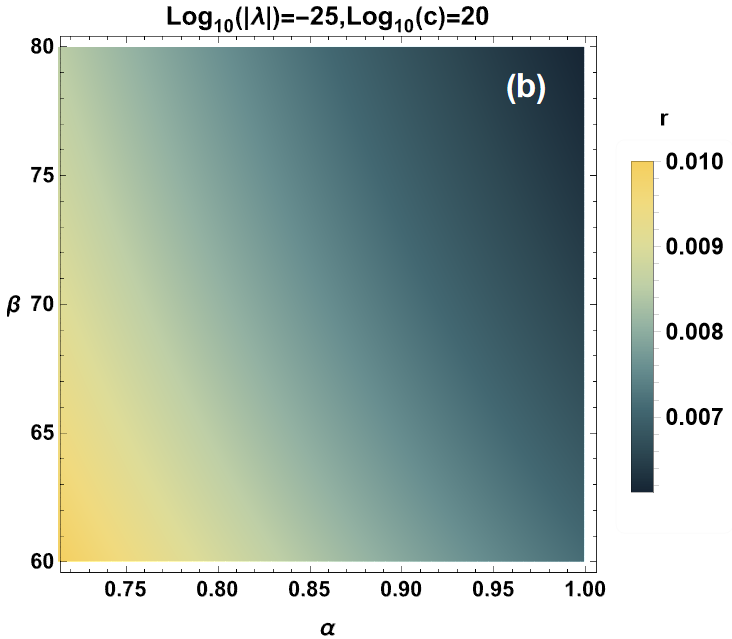}\\
 \includegraphics[width=8cm, height=7cm]{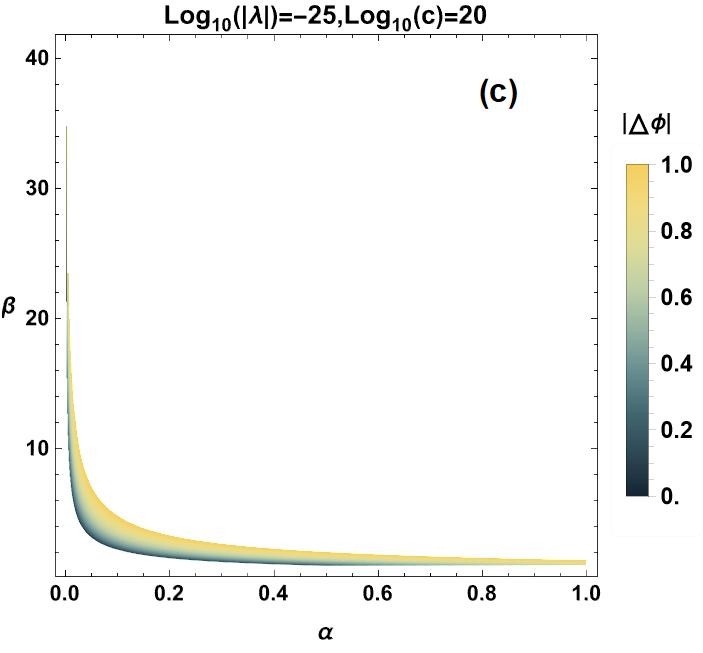}
 \hspace{0.07cm}
 \includegraphics[width=7.9cm, height=7cm]{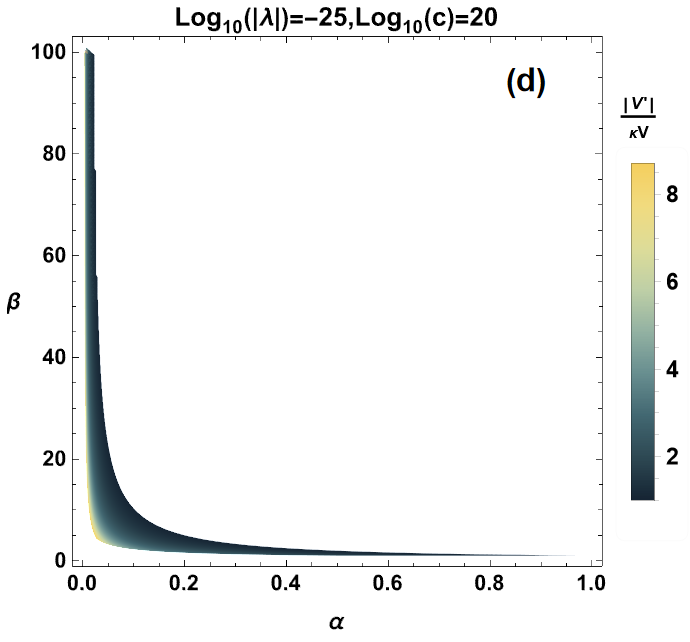}
  \end{tabbing}
\caption{ (a): Scalar index $n_s$ in  terms  of the free parameter $\beta$ and scaling parameter $\alpha$. (b):  Scalar to tensor ratio $r$ in  terms of  $\beta$ and $\alpha$.  (c):  Distance conjecture compatible values $\Delta\phi$ in  terms  of $\beta$ and $\alpha$.
(d): De Sitter conjecture bound runway instability values  $\frac{\vert V^\prime\vert}{\kappa V}$ in  terms  of  $\beta$ and $\alpha$.}
\label{F1}
\end{center}
\end{figure}

The first selected model concerns $n=2$.  To extract the numerical  values of the relevant quantities,  we consider  the moduli space point $P_1\equiv(\lambda=-10^{-25},c=10^{20},N=54,\beta=70.1,\alpha=0.9)$. The associated scalar spectral index is found to be  $n_S=0.961971164$, while the scalar to tensor ratio  is $r=0.00722608669$ represented by a dark Tile in Fig.\ref{F1}.(a) and Fig.\ref{F1}.(b), respectively.  Moreover, we find that the tensor spectral index is $n_\tau=-9.03260836\ 10^{-4}$. These results are well bound by the observational data \cite{IST6}. Moreover, they are  consistent with the stringy constraint $r\approx -\Delta n_\tau$ \cite{2,17}, where  one has $\Delta=(8+64\frac{H}{\alpha n_\tau}\xi^\prime)=8$. This can be owed to the fact that the coupling function is not normalizable.   Concerning the swampland program,  we  get   the numerical value of the field distance being  $\Delta\phi=8.64684582$  where  we have found  field values   $\phi_i\approx 0.999345601$ and $\phi_f=9.64619160$.  It has been remarked that  this value asserts that the field distance does not adhere to the distance conjecture, where all the adherent values are represented in   Fig.\ref{F1}.(c). For the runway instability  and the tachyonic instability,   we  find  $\frac{\vert V^\prime\vert}{\kappa V}=0.0316800065$ and $-\frac{V^{\prime\prime}}{\kappa^2V}=-0.00100362289$, respectively. They are in disagreement  with the de Sitter conjecture, as seen in Fig.\ref{F1}.(d) representing the runaway instability. This approach seems to have the characteristics associated  with  the swampland treatment of inflation mentioned in \cite{R11}, owing the disagreement with the de Sitter conjecture due to the small values of the slow roll indices  $\epsilon_1=4.51630418\ 10^{-4}$ and $\epsilon_2=0.0181111638$. Even if a region of the moduli space is compatible with all the criteria, it must be  a highly finely tuned as seen in  Fig.\ref{F1}.(a), where the Planck data bound region is represented as an extremely thin brushstroke. The remaining  indices have the following values $\epsilon_3=8.70766866\ 10^{-42}$ and $\epsilon_4=-3.65120831\ 10^{-23}$. We also highlight the value of the primordial sound wave $c_A=1.0$ and the value of the scalar potential $V(\phi_i)=9.79108165\ 10^{-21}$.

It is worth noting that the  tachyonic instability values  are not   approached  due to the  absence of   the de Sitter conjecture bound values in the  selected free parameter ranges.\\


\subsection{Models with $n\geq3$}
Here,   we consider  certain  specific models  with $n\geq3$.  Precisely, we study the  models  with $n=3$, $n=4$, and $n=5$  by considering  different points in the moduli  space.  Instead of giving graphical discussions, we only provide the  numerical values of the involved quantities   being   summarized
in the Table 1.\\

\begin{table}[!h]
\begin{center}
\begin{tabular}{|c|c|c|c|}
\hline
 &    $n=3$ for the point &   $n=4$  for the point &   $n=5$ for the point \\
 & \small{${\scriptscriptstyle(\lambda=-10^{-25},c=10^{20},N=50,\beta=40,\alpha=0.9)}$} & \small{${\scriptscriptstyle(\lambda=-10^{-25},c=10^{20},N=50,\beta=50.2,\alpha=0.9)}$} & \small{${\scriptscriptstyle(\lambda=-10^{-30},c=10^{20},N=50,\beta=52.7,\alpha=0.9)}$} \\
\hline
 $n_{S}$ &  $0.961672127$ & $0.966385067$ & $0.966511965$\\
\hline
$r$ &  $0.00888221804$ & $0.00762708206$ & $0.00686550513 $ \\
\hline
$n_{\tau}$ &  $-0.0011027726$ & $-0.00953385257$ & $-0.00858188141$\\
\hline
$\vert \Delta\phi\vert$ &  $3.17418861$ & $2.02584505$ & $1.47910070$\\
\hline
$\frac{\vert V\vert}{\kappa V}$ &  $0.0325471498$ & $0.0340702087$ & $0.0308794845$\\
\hline
$-\frac{V^{\prime\prime}}{\kappa^2V}$ &  $-0.00105931703$ & $-0.00116077904$ & $-0.00953542476$\\
\hline
\end{tabular}
\caption{ Numerical values of the models $n=3$, $n=4$ and $n=5$.}
\label{Tab1}

\end{center}

\end{table}
Taking in account the numerical approach,  we  should take the showcased values with a grain of salt, where it has been  observed that the value of $\phi_i$ being a  solution of Eq.(\ref{Eq.3.4}) has been  found with an error of the order $\mathcal{O}(10^{-2})$. Nonetheless, the characteristics of the models are not effected by such  an error.\\
For the selected values of $n$, it  follows from the table that the scalar spectral index,   the scalar to tensor ratio and the tensor spectral index  values match  perfectly with the observational data.  Concerning, the stringy constraint, one can calculate  $\delta=\vert\Delta-\frac{r}{\vert n_\tau\vert}\vert$ for different values of $n$. Taking   $n=3$, we find  $\delta=0$. For $n=4$ and $n=5$, however,  we obtain $\delta=2.403\ 10^{-4}$ and $\delta=3.81\ 10^{-6}$,  respectively. It is observed that  the discrepancy with the stringy constraint grows by increasing $n$. This can be compensated by taking very small values of the coupling constant $\lambda$, rendering a conventional inflationary theory. Such small values do not reflect on the order of the differential forms of the coupling functions. At the chosen moduli space  point, for $n=3$,   we have  found $\xi^\prime=-9.22\ 10^{-4}$ and $\xi^{\prime\prime}=-0.144844$. However, the choice of the small value of the coupling constant could  insure the consistency with the stringy constraint $r\approx-\Delta n_\tau$. This characteristic is a consequence of the non-normalizable coupling function, where the naturalness argument appears to have effect on the relevant quantities.\\
Moreover, a close  inspection shows   that the field distance decreases by  increasing $n$. For $\mathcal{O}(10)$ order values of $\beta\sqrt{\alpha}$, we  obtain   $\phi_f\approx 1$ and $\vert\Delta\phi\vert< 1$ satisfying the distance conjecture.\\

Regarding the de Sitter conjecture, we remark  that it is not met for the selected values of $n$. In fact, it  has been  assumed that all the possible models in this frame do not adhere to the said conjecture. This discrepancy is a characteristic of such an  inflationary theory, as pointed in \cite{R11}. In an epistemological approach, such an effort concerns the falsification aspects of the theory, where the absence of  such an aspect indicates a very large string-landscape/moduli-space of the effective field theory. 
This issue is mediated in \cite{RG2}, by implementing normalizable coupling functions,  since the naturalness of the effective field theory  seems to have effects on the relevant quantities. 
\section{Conclusion}
In this paper,  we have investigated  the inflationary hypergeometric behaviors in the swampland program via the falsification scenario.  In particular, we have exposed   the swampland criteria for  GB stringy   corrections in the    rescaled gravity within a linear function  of   the  Ricci scalar  namely $F(R)=\alpha R$.   Using special functions,   we  have  proposed   and examined   inflationary  models from hypergeometric scalar coupling scenarios.   Concretely,   we  have provided  generalized exponential potentials from  such hypergeometric behaviors in the   swampland program. Precisely, we have      considered  selected  models with different values of the parameter $n$  by means of   the slow-roll mechanism.  For $n=2$, we have  given   numerical and  graphical investigations by computing the relevant quantities which are found to be compatible with the  empirical Planck data.  The swampland aspect seems to have the characteristics postulated for conventional inflationary theories.  
 For   $n=3$, $n=4$ and $n=5$, however, we  summarized  the obtained values  in Table 1.   Analyzing this table, we  have observed the  compatibility of the Planck observational  data. Via  a numerical manner,  we  have shed light  on the discussion provided in \cite{R11}, regarding the discrepancy of the swampland criteria with conventional inflationary models, namely the de Sitter conjecture.

This works comes up with certain open issues.  One of them  is  to consider extra coupling scenarios inspired by string theory corrections which could  bring  complete differential equations associated with special functions. Other questions could  arise including  the extant of the naturalness argument effects on the relevant quantities of the effective field theory in connection  with swampland program.   These questions could be addressed  elsewhere. 



\end{document}